# Measurement of the Background Activities of a $^{100}$Mo-enriched Powder Sample for an AMoRE Crystal Material by Using a Single High-Purity Germanium Detector

Su-yeon Park

*Department of Physics, Ewha Womans University, Seoul 03760, Korea*

Insik Hahn

*Department of Science Education, Ewha Womans University, Seoul 03760, Korea, and*

*Center for Exotic Nuclear Studies, Institute for Basic Science (IBS), Daejeon 34126, Korea*

Woon Gu Kang, Gowoon Kim*, Eunkyung Lee and Douglas S. Leonard

*Center for Underground Physics, Institute for Basic Science (IBS), Daejeon 34126, Korea*

Vladimir Kazalov

*Baksan Neutrino Observatory, Institute for Nuclear Research of the Russian Academy of Science, Kabardino-Balkaria 361609, Russia*

Yeong Duk Kim

*Center for Underground Physics, Institute for Basic Science (IBS), Daejeon 34126, Korea,*

*Department of Physics and Astronomy, Sejong University, Seoul 05006, Korea, and*

*Institute for Basic Science (IBS) School, University of Science and Technology (UST), Daejeon 34113, Korea*

Moo Hyun Lee

*Center for Underground Physics, Institute for Basic Science (IBS), Daejeon 34126, Korea, and*

*Institute for Basic Science (IBS) School, University of Science and Technology (UST), Daejeon 34113, Korea*

Elena Sala

*Center for Axion and Precision Physics Research, Institute for Basic Science (IBS), Daejeon 34051, Korea*

The Advanced Molybdenum-based Rare process Experiment (AMoRE) searches for neutrino-less double-beta (0νββ) decay of $^{100}$Mo in enriched molybdate crystals. The AMoRE crystals must have low levels of radioactive contamination to achieve low background signals with energies near the Q-value of the $^{100}$Mo 0νββ decay. To produce low-activity crystals, radioactive contaminants in the raw materials used to form the crystals must be controlled and quantified. $^{100Enr}$MoO$_3$ powder, which is enriched in the $^{100}$Mo isotope, is of particular interest as it is the source of $^{100}$Mo in the crystals. A high-purity germanium detector having 100% relative efficiency, named CC1, is being operated in the Yangyang underground laboratory. Using CC1, we collected a gamma spectrum from a 1.6-kg $^{100Enr}$MoO$_3$ powder sample enriched to 96.4% in $^{100}$Mo. Activities were analyzed for the isotopes $^{228}$Ac, $^{228}$Th, $^{226}$Ra, and $^{40}$K. They are long-lived naturally occurring isotopes that can produce background signals in the region of interest for AMoRE. Activities of both $^{228}$Ac and $^{228}$Th were < 1.0 mBq/kg at 90% confidence level (C.L.). The activity of $^{226}$Ra was measured to be 5.1 ± 0.4 (stat) ± 2.2 (syst) mBq/kg. The $^{40}$K activity was found as < 16.4 mBq/kg at 90% C.L.

* E-mail: kkw-owo@hanmail.net (Gowoon Kim)

Fax: +82-42-878-8509

# I. INTRODUCTION

Searches for neutrinoless double-beta (0νββ) decay in isotopes such as $^{100}$Mo can potentially determine if the neutrino is a Majorana particle or not. If the $^{100}$Mo 0νββ decay occurs, the Q-value energy of 3034.4 keV [1] will be carried away by the two emitted electrons, creating an observable peak in the energy spectrum of a detector. The half-life of the decay is estimated to be longer than $1.1 \times 10^{24}$ years [2]. To reach sensitivities of the 0νββ decay at or above this half-life, the rate of background events in the region of interest (ROI) must be less than $10^{-4}$ count/keV/kg/year [3].

The Advanced Molybdenum-based Rare process Experiment (AMoRE) searches for neutrinoless double-beta (0νββ) decay of $^{100}$Mo by using enriched molybdate crystals. Currently, the full experimental setup, AMoRE-II, is being prepared to observe the decay in detectors constructed with cryogenic crystals containing molybdenum enriched in the $^{100}$Mo isotope [4, 5]. Around 200 kg of molybdate crystals will be in operation.

The AMoRE detectors have full width at half maximum (FWHM) energy resolutions of about 10 keV [6], implying ROI around the Q-value from 3024.4 keV to 3044.4 keV, if the ROI width is defined as twice the FWHM. Daughter isotopes of the $^{232}$Th and the $^{238}$U decay chains can produce background events in the AMoRE in this energy range via multiple mechanisms involving single or coincident emissions. In particular, as explained in Ref. 7, the $^{212}$Bi-$^{212}$Po and the $^{214}$Bi-$^{214}$Po decay sequences can produce backgrounds in the ROI. Coincident emissions from $^{208}$Tl decay, primarily from 583.2 keV and 2614.5 keV gammas, can also produce signals in the AMoRE ROI.

Two-neutrino double-beta decay of $^{100}$Mo can produce electrons with energies up to the Q-value. Sufficiently high-activity decays from radioactive contaminants can potentially create background signals in the ROI through random coincidence between two such decays or between one contaminant decay and a two-neutrino decay. For example, the most energetic decay mode of $^{40}$K has a Q-value of 1504.4 keV, well below the AMoRE ROI. However, if the activity of $^{40}$K in the crystals is high enough, a signal in the ROI can still result via random coincidence with other activities, including two-neutrino decay.

Internal background levels of $^{40}$Ca$^{100}$MoO$_4$ crystals for the AMoRE were measured to obtain activities of $^{214}$Po from the $^{238}$U decay chain and $^{216}$Po from the $^{232}$Th decay chain [8-10]. The AMoRE detector components were also scanned to acquire activities of background contaminants [11, 12]. Based on these data, the $^{232}$Th and the $^{238}$U decay chains and $^{40}$K decays were simulated to find which parts of the AMoRE detector system contribute significant background signals in the ROI [13]. Previous measurements and simulations reported in Refs. 8-13 showed that internal radioactive sources in the crystals likely contribute most to the background in the ROI.

As a way to reduce the levels of background contaminants in the crystals, the $^{100}$Mo-enriched powder is purified by using physical and chemical methods before being used for crystal growth [14]. Monitoring the quality of purification before proceeding to crystal growth is essential. To understand the starting point for this process, we measured the activities of contaminants in an unpurified powder sample.

A high-purity germanium (HPGe) detector named CC1 with a 100% relative efficiency is located in the Yangyang underground laboratory (Y2L). The underground location, at a depth of 700 m, is used to reduce the background from cosmic radiation [11, 12, 15]. Lead and copper blocks shield CC1. The background level of CC1 is around 0.0078 s$^{-1}$ in the energy range from 50 keV to 4000 keV [15].

The levels of radioactive contamination in a 1.6-kg sample of powder were measured using CC1. We analyzed activities of several radioactive isotopes, including $^{228}$Ac, $^{228}$Th, $^{226}$Ra, and $^{40}$K, which can cause background signals in the ROI of AMoRE.

## II. EXPERIMENTAL SETUP

The germanium crystal of the CC1 detector has a diameter of 81.4 mm and a height of 81.7 mm. The crystal is in an aluminum canister that has an outer diameter of 105 mm and a height of 120 mm. A sample space surrounds the canister. As described in Ref. 15, the detector and sample space are surrounded by a layered shield. The shielding consists of, from inside out, a 50-mm-thick layer of lead, a 100-mm-thick layer of copper, and an outer lead shield. The vertical walls of the outermost shield are 150-mm thick while the upper and the lower layers are 100-mm thick. Figure 1 shows a schematic of the CC1 detection system, including the geometry of the powder sample.

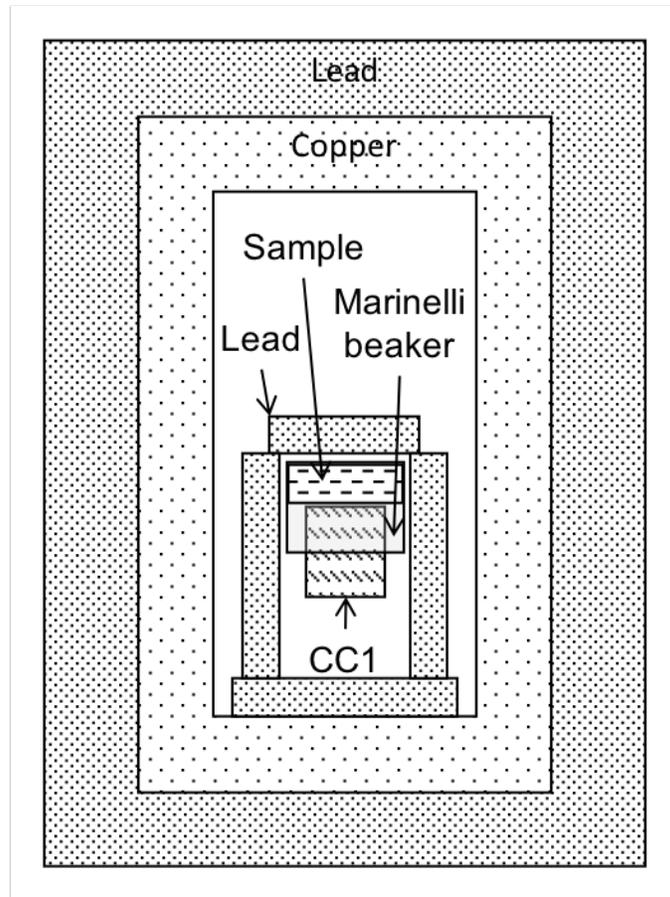

Fig. 1. Schematic of the experimental setup with the CC1 detector system, lead and copper shields, and the powder sample in a Marinelli beaker.

A total of 10 kg of $^{100}$Mo-enriched powder was provided by JSC Isotope [16]. The supplier certified that it contained 6.8 kg of molybdenum and that the enrichment level of $^{100}$Mo was 96.4%. The $^{100}$Mo-enriched powder was divided and packaged into 15 separate plastic bags. Three bags, which weigh 1.6 kg in total, could fit in a Marinelli beaker. The shape and the position of the powder sample on CC1 could be fixed during data taking. The Marinelli beaker containing the powder sample was placed around CC1, as shown in Figs. 1 and 2.

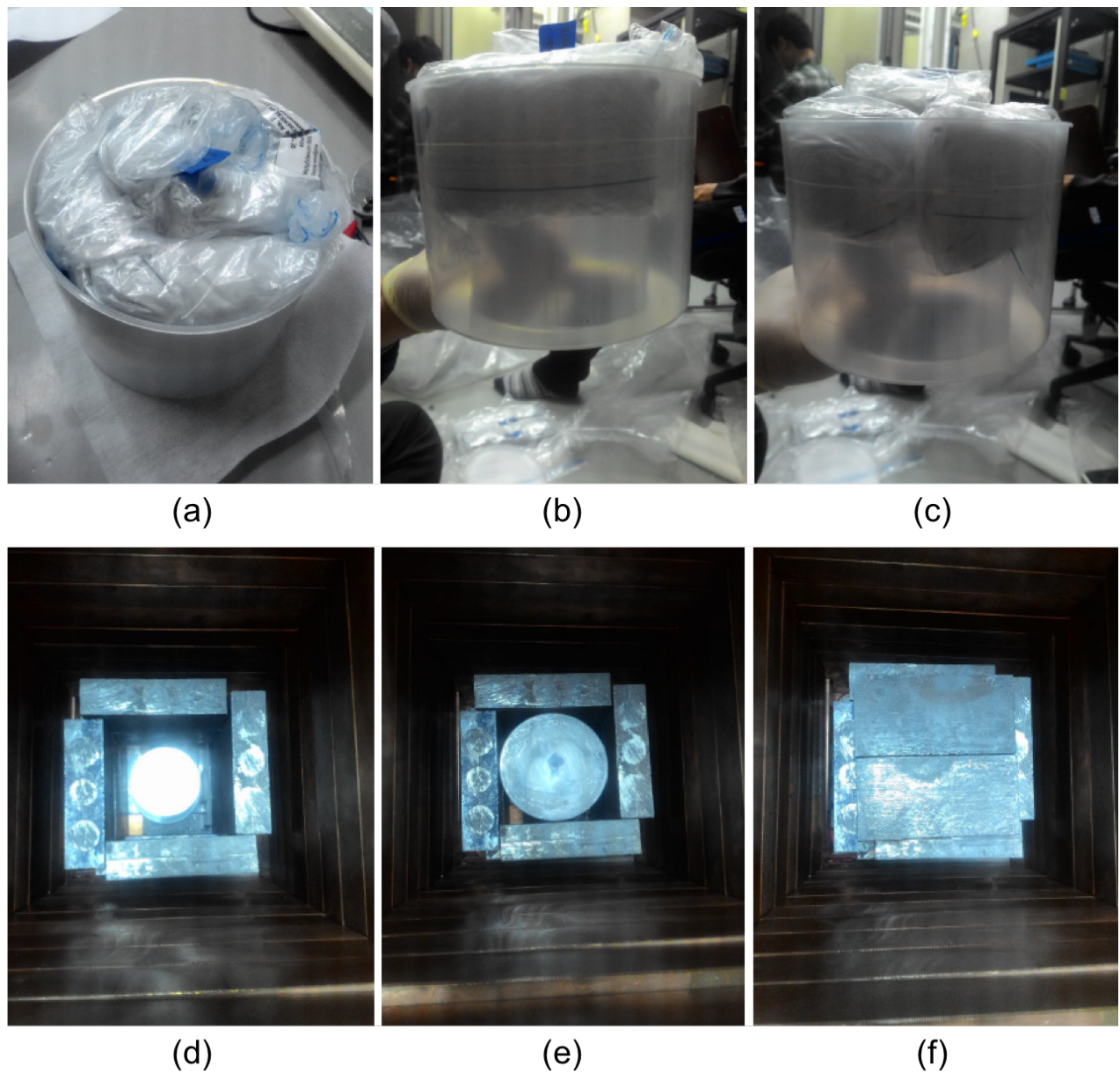

Fig. 2. (a-c) Powder sample in a Marinelli beaker and (d-f) its installation on CC1.

Signals were sent from the pre-amplifier to a shaping amplifier configured with a 6-μs shaping time. Signals from the outputs of the shaping amplifier were digitized by a multichannel analyzer and stored to disk for later analysis. The data were collected for 340 h.

$^{222}$Rn gas entering the detection volume from laboratory air can produce background signals for the $^{226}$Ra sub-chain. The detection volume was continuously flushed with boil-off nitrogen gas from a liquid-nitrogen bottle in order to reduce this effect.

In order to subtract signals arising from contaminants in the detector system itself, we obtained background data with the sample space empty, i.e., without the presence of the powder sample. The background data and the powder data were taken in the same experimental environment and conditions, with, for example, a similar nitrogen gas flushing rate. The total live-time for background data was 807 h.

### III. ANALYSIS

All of the radioactive decay information used in this analysis were from the National Nuclear Data Center (NNDC) [17]. The detection efficiencies were calculated using GEANT4 Monte Carlo simulations [18].

The energy spectra collected with the powder sample were calibrated using the most probable centroid values from fits to gamma peaks at 186.2 keV from $^{226}$Ra decay, 295.2 keV and 352.0 keV from $^{214}$Pb, 609.3 keV, 1120.3 keV, and 1764.5 keV from $^{214}$Bi, and 1460.8 keV from $^{40}$K. Each peak was fitted to a Gaussian distribution function added to a parameterization of the continuum background caused by Compton scattering. A quadratic function was used to find a relation between ADC channels and energies, as shown in Fig. 3(a). The calibration function determined for the powder data was

$$Energy\ [keV] = -0.61 + 0.40 \times ADC + 2.14 \times 10^{-8} \times ADC^2,$$

where *ADC* means the ADC channel number.

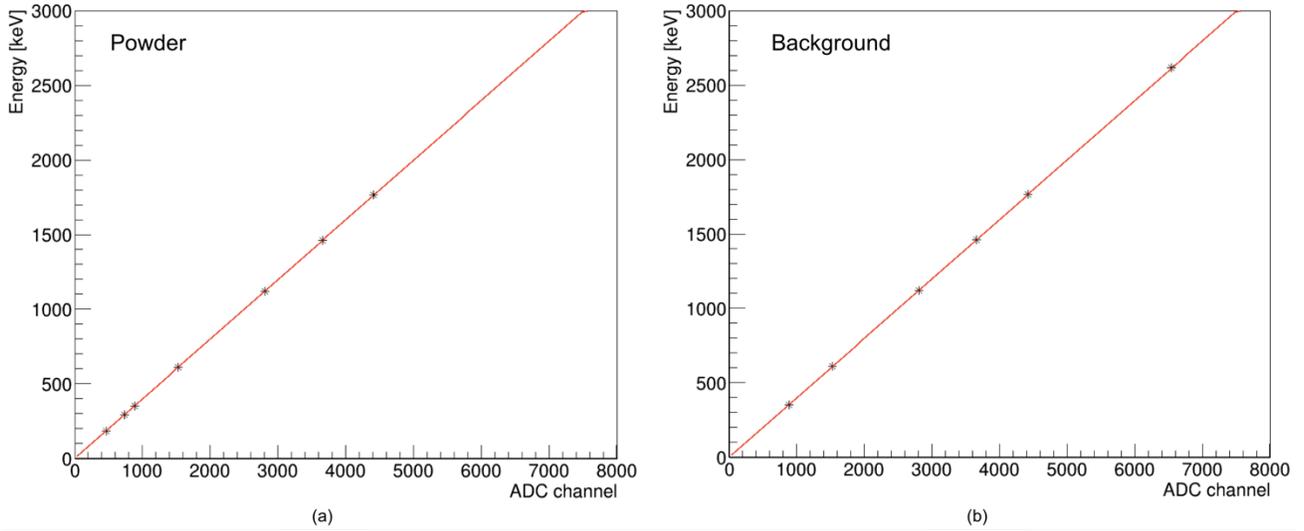

Fig. 3. Energy vs. ADC channel with calibration peak points and fitted functions: (a) powder data and (b) background data. Quadratic functions were used for calibration fits.

The energies of the observed $^{214}$Bi peaks at 609.3 keV and 1120.3 keV were inferred from the calibration and compared to the known values to confirm the uncertainty of the energy calibrations. The peak positions put into the calibration function were obtained from fits in the same way as for the calibration peaks. The inferred energies were lower than the known values by 0.2 keV, a negligible difference. The energy resolutions were 1.5 keV (FWHM) at both 609.3 keV and 1120.3 keV.

The background data were calibrated with the 352.0 keV, 609.3 keV, 1120.3 keV, 1460.8 keV, 1764.5 keV, and 2614.5 keV peaks in the same way. A quadratic function was also used, as shown in Fig. 3(b). The calibration function for the background data was

$$Energy\ [keV] = -0.92 + 0.40 \times ADC + 3.41 \times 10^{-8} \times ADC^2.$$

The observed positions of 609.3 keV and 1460.8 keV peaks were confirmed in the same way as for the powder data, and differences were 0.2 keV and 0.1 keV lower than the known values, respectively.

Figure 4 shows the energy spectra of powder data and background data. Energy spectra were divided into six 500-keV regions. Most of the marked peaks in Fig. 4 were produced by gamma emissions from decays of $^{228}$Ac, $^{228}$Th, and $^{226}$Ra. The 1460.8-keV peak from $^{40}$K was also observed. Activities of these radioisotopes, which can cause signals in the AMoRE ROI, were analyzed.

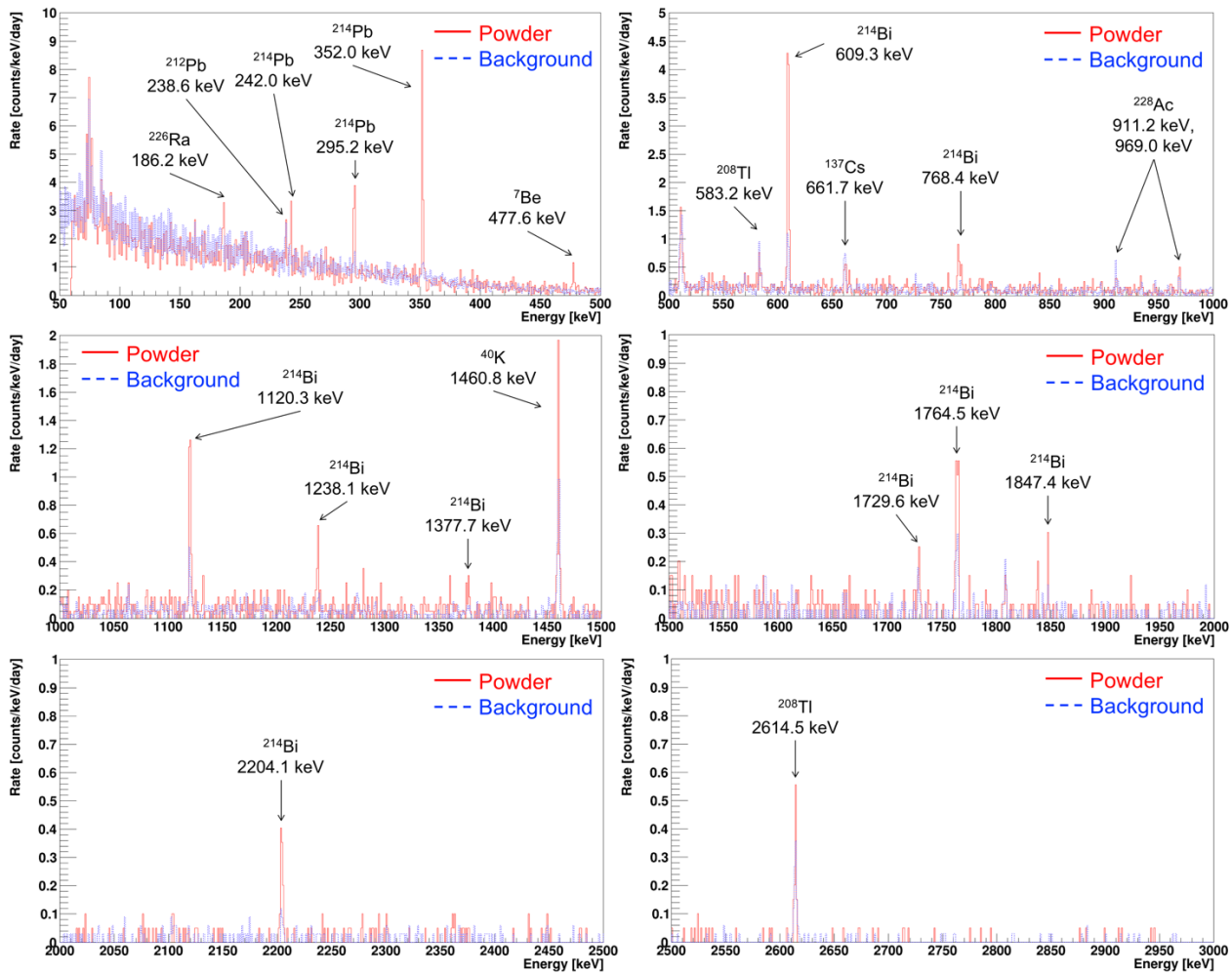

Fig. 4. Energy spectra of the powder data (red) and the background data (blue). Energy spectra were divided into six 500-keV regions. The gamma peaks were indicated with the energies and their mother isotopes.

The 477.6 keV peak from the electron capture of $^7$Be was also observed. Reactions between secondary cosmic rays and nitrogen or oxygen can produce this isotope [19]. The $^{100}$Mo-enriched powder was delivered via a few flights from a production site in Russia to Y2L in Korea. We assume that the $^7$Be in the powder sample was produced as a consequence of the $^{100}$Mo-enriched powder's exposure to cosmic rays during the flights. However, the half-life of $^7$Be is 53.2 days, which is short compared to the AMoRE-II preparation period, and its Q-value of 861.8 keV is far lower than the AMoRE ROI centered around 3034.4 keV. Therefore, the activity of $^7$Be was not analyzed.

**1. Background Activities from Labels**

Because of logistical constraints and to avoid contaminating the powder sample or losing any of the valuable material, it was kept in the original plastic bags and placed in a Marinelli beaker. All the bags had labels attached outside, as shown in Fig. 2. When preparing the powder sample, the labels were kept on the bags to avoid damaging or tearing the bags. Three pieces of labels were thus installed together with the powder sample in CC1. The labels were composed of paper, adhesive, and ink, which can both contain relatively high levels of radioactive contaminants. Contributions from these activities should be subtracted from the observed count rates. When the powder sample was later processed for purification at a separate facility, the labels were retrieved. Several samples of labels were then assayed using CC1. Figure 5 shows some of the label samples. The activities obtained from these assays, in mBq/label, are summarized in Table 1.

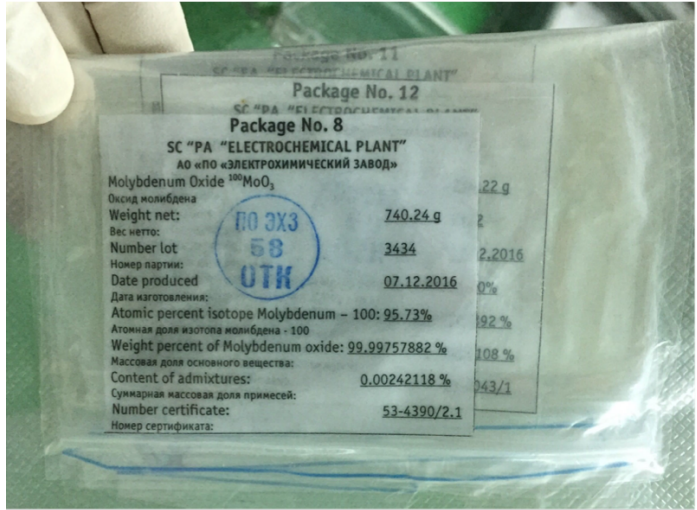 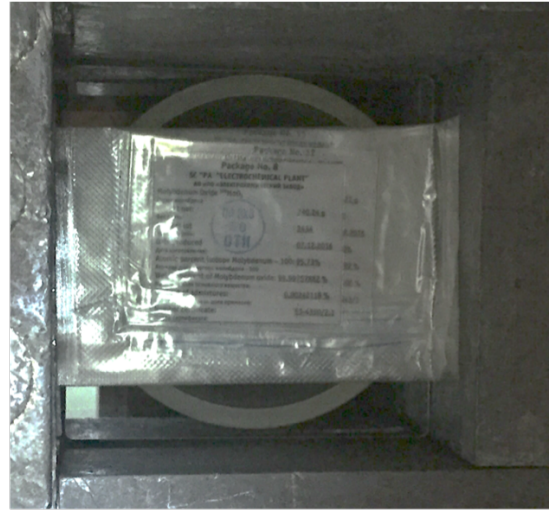

(a)    (b)

Fig. 5. For subtracting the contributed counts of gammas from the labels in the powder data, we measured the activities in the label samples on CC1. (a) Labels in plastic bags are vacuum sealed and (b) the samples are placed on CC1 for the activity measurement.

Table 1. Measured activities of various isotopes in the representative label samples in unit of mBq/label. Three labels in total were installed together with the powder sample.

| Isotope | $^{228}$Ac | $^{228}$Th | $^{226}$Ra | $^{40}$K |
|---|---|---|---|---|
| Activity [mBq/label] | 0.06 ± 0.03 (< 3.8 at 90% C.L.) | 0.07 ± 0.02 | 2.3 ± 0.1 | 0.9 ± 0.2 |

The measured label activity values were used to estimate the contributions by the label contaminations to the peak rates observed in the powder activity measurement. Uncertainties from these contributions were also considered. For a particular peak from a particular activity, the number of counts $C_L$ contributed by the labels during the measurement time $T_P$ of the powder sample was determined as

$$\frac{C_L}{T_P} = A_L \times \varepsilon_L \times N_L \times B.R. \times G.I.,$$

where $A_L$ is the observed specific activity of the labels in Table 1, $\varepsilon_L$ is the full-energy detection efficiency for a gamma of interest produced from random locations within the labels during the powder activity measurement, $N_L$ is the number of labels included in the powder activity measurement, *B.R.,* is the branching ratio for the isotope in question, and *G.I.* is the gamma intensity.

## 2. Detection Efficiency and Simulation

The decay chains of $^{232}$Th and $^{238}$U, as well as decays of $^{40}$K, were simulated in the powder and labels using the GEANT4 simulation toolkit to determine detection efficiencies. Full decays were simulated including coincident gamma emission. The geometries of the CC1 detector system, the powder sample, the labels, and the Marinelli beaker were included in the simulation. Details are shown in Fig 6.

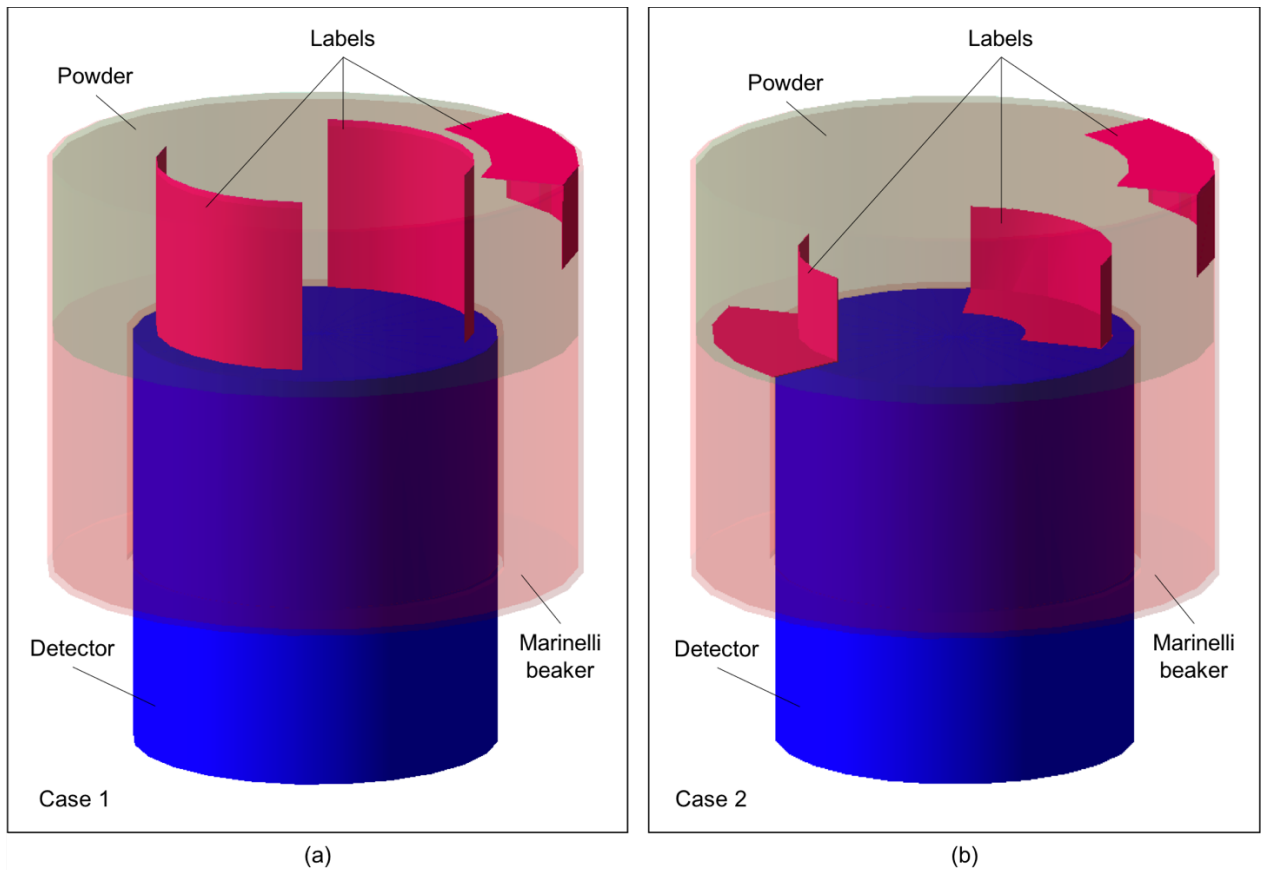

Fig. 6. Two different simulated geometries for the powder sample and the labels in a Marinelli beaker on CC1: (a) the geometry of case 1, which gave the minimum values of the detection efficiency for the label contaminants and (b) that of case 2, which produced the maximum values. The blue object is the cryostat for the CC1 detector. The Marinelli beaker is shown in pink. Transparent dark green represents the powder sample. The three red items are the labels.

The detection efficiency, $\varepsilon$, of a gamma with emitted energy $E$ is calculated as

$$\varepsilon = \frac{N_d}{N_g},$$

where $N_d$ is the area of the Gaussian distribution, in units of counts, determined from the fit to the full-energy peak at $E$, and $N_g$ is the number of gammas generated with energy $E$. Gammas were generated in random locations within the geometry of the powder sample. For the purpose of subtracting contributions from the labels, the simulation of decays in the labels was generated separately.

Because the positions and the shapes of the labels were not fully recorded, a few sample-installation cases that represent the extreme possibilities for label detection efficiency were simulated. From the photos in Fig. 2, one label faced outward and was folded in half. The exact orientations of the other labels were not known. Through a combination of straight and folded labels, four representative geometrical cases were considered. All of them were simulated, and the two cases that gave the minimum and the maximum detection efficiencies, respectively, are shown in Fig. 6. Table 2 lists detection efficiencies of the powder sample, $\varepsilon_P$, and the labels, $\varepsilon_L$.

Table 2. Simulated detection efficiencies $\varepsilon_P$ and $\varepsilon_L$ for the powder sample and the labels, respectively, for two simulation cases representing the extremes in the possible label positions.

| Isotope | | | Energy [keV] | Detection efficiency (%) | | | |
|---|---|---|---|---|---|---|---|
| | | | | Case 1 | | Case 2 | |
| | | | | $\varepsilon_P$ | $\varepsilon_L$ | $\varepsilon_P$ | $\varepsilon_L$ |
| $^{228}$Ac | | | 911.2 | 1.7 | 1.2 | 1.6 | 2.4 |
| | | | 969.0 | 1.6 | 1.2 | 1.6 | 2.3 |
| $^{228}$Th | $^{212}$Pb | | 238.6 | 2.8 | 2.0 | 2.8 | 5.1 |
| | $^{208}$Tl | | 583.2 | 1.9 | 1.4 | 1.9 | 2.6 |
| | | | 2614.5 | 0.8 | 0.7 | 0.8 | 1.0 |
| $^{226}$Ra | $^{214}$Pb | | 295.2 | 3.1 | 2.2 | 3.1 | 5.3 |
| | | | 352.0 | 2.8 | 2.0 | 2.8 | 4.4 |
| | $^{214}$Bi | | 609.3 | 1.9 | 1.5 | 1.9 | 2.7 |
| | | | 1120.3 | 1.5 | 1.2 | 1.5 | 2.0 |
| | | | 1764.5 | 1.3 | 1.0 | 1.3 | 1.8 |
| $^{40}$K | | | 1460.8 | 1.4 | 1.1 | 1.3 | 2.0 |

Dead layers, inactive volumes in a germanium crystal, act like shielding layers and reduce the detection volume of the germanium [20]. For this reason, the efficiency was calibrated with a multi-nuclide source with known activities. The simulation geometry of the CC1 detector system was fine-tuned to produce efficiencies matching the multi-nuclide source data. After tuning the simulation, the discrepancy between the calibration data and the simulation was found to be less than 10% [12]. For simulation studies of similar measurements, we have found that sample geometry errors contribute at most a few percent. We include a 10% systematic error to account for calibration discrepancies and sample geometry errors.

3. Count and Activity

A parent activity $A_P$ from the powder sample derived from any single gamma peak in one of the decay-chain isotopes is given by

$$A_P = \frac{\frac{C_O}{T_P} - \frac{C_L}{T_P} - \frac{C_B}{T_B}}{\varepsilon_P \times M_P \times B.R. \times G.I.},$$

where $C_O$ is peak counts observed in the powder data, $C_L$ is the estimated contribution to $C_O$ from contaminations in the labels, $C_B$ is peak counts observed in the background data, $T_B$ is the measurement time of the background data, and $M_P$ is the mass of the powder sample. The background activities in the powder sample were analyzed with the two maximally different detection efficiency scenarios described above, and the results for both cases were obtained. Two isotopes of $^{228}$Th and $^{226}$Ra were assumed to be in equilibrium with their respective decay chains, a good assumption given the short half-lives of the decay-chain daughters. With this assumption, all activities in the $^{228}$Th or $^{226}$Ra sub-chains were, with appropriate branching ratios applied, assumed to be valid measures of the respective parent activities. Results from the same chain were thus averaged together. Differences between the minimum and the maximum activity results were used for an estimate of geometrical uncertainty in the contribution from the label contaminants. Table 3 lists the counts in the gamma peaks from $^{228}$Ac, $^{228}$Th, $^{226}$Ra, and $^{40}$K for the powder data and the background data. Table 4 lists the analysis results using the two most extreme simulation cases.

Table 3. Peak rates for the activity analysis. For each gamma peak, $C_O$ is the total observed number of counts in the full-energy peak from the powder data, $C_L$ is the estimated contribution to $C_O$ from contaminations in the labels, and $C_B$ is counts from the background peak intrinsic to the detector's construction. For comparison, counts are normalized to the acquisition times $T_P$ (340 h) and $T_B$ (807 h) for the powder and the background data, respectively.

| Isotope | | | Energy [keV] | Rate [× 10$^{-2}$ counts/hour] | | | |
|---|---|---|---|---|---|---|---|
| | | | | $C_O/T_P$ | Case 1 $C_L/T_P$ | Case 2 $C_L/T_P$ | $C_B/T_B$ |
| $^{228}$Ac | | | 911.2 | 2.9 ± 1.3 | 0.21 ± 0.09 | 0.40 ± 0.17 | 3.5 ± 0.8 |
| | | | 969.0 | 4.4 ± 1.5 | 0.13 ± 0.05 | 0.23 ± 0.10 | 2.3 ± 0.7 |
| $^{228}$Th | $^{212}$Pb | | 238.6 | 11.4 ± 3.1 | 0.64 ± 0.14 | 1.62 ± 0.36 | 10.7 ± 1.1 |
| | $^{208}$Tl | | 583.2 | 5.2 ± 1.9 | 0.32 ± 0.07 | 0.59 ± 0.13 | 7.6 ± 1.1 |
| | | | 2614.5 | 6.2 ± 1.3 | 0.18 ± 0.04 | 0.27 ± 0.06 | 3.8 ± 0.7 |
| $^{226}$Ra | $^{214}$Pb | | 295.2 | 37.2 ± 2.5 | 10.2 ± 0.5 | 24.4 ± 1.3 | 3.0 ± 0.9 |
| | | | 352.0 | 70.4 ± 3.0 | 17.6 ± 0.9 | 39.5 ± 2.1 | 4.8 ± 0.7 |
| | $^{214}$Bi | | 609.3 | 53.9 ± 2.6 | 16.8 ± 0.9 | 30.3 ± 1.6 | 8.9 ± 1.2 |
| | | | 1120.3 | 16.7 ± 2.5 | 4.4 ± 0.2 | 7.4 ± 0.4 | 3.4 ± 0.8 |

|   |   | 1764.5 | 10.3 ± 1.8 | 3.8 ± 0.2 | 6.7 ± 0.4 | 2.8 ± 0.7 |
|---|---|---|---|---|---|---|
|   | $^{40}$K | 1460.8 | 19.6 ± 2.6 | 1.1 ± 0.2 | 1.9 ± 0.4 | 10.5 ± 1.2 |

Table 4. Activities of various isotopes in the powder sample derived using two different simulations for the label contamination efficiencies based on the most extreme possible arrangements of the labels.

| Isotope | Activity ± statistical uncertainty [mBq/kg] | |
|---|---|---|
|   | Case 1 | Case 2 |
| $^{228}$Ac | 0.06 ± 0.52 | 0.001 ± 0.535 |
| $^{228}$Th | 0.3 ± 0.4 | 0.2 ± 0.4 |
| $^{226}$Ra | 6.9 ± 0.3 | 3.4 ± 0.4 |
| $^{40}$K | 9.5 ± 3.3 | 8.6 ± 3.4 |

### IV. RESULTS AND CONCLUSIONS

Table 5 summarizes the background activity results of the powder sample. The observed activities for $^{228}$Ac, $^{228}$Th, and $^{40}$K, are found to be within 3σ of zero and are reported as upper limits. The limits were calculated conservatively using the label geometry assumption that resulted in the highest activities. This assumption is represented by the case 1 activities shown in Table 4. An additional systematic uncertainty of 10% is attributed to other efficiency simulation errors. The activities of both $^{228}$Ac and $^{228}$Th were < 1.0 mBq/kg at 90% C.L. The activity of $^{40}$K was < 16.4 mBq/kg at 90% C.L.

Table 5. Results of activity analyses for the isotopes that can make background signals in the AMoRE-II ROI. Limits at 90% C.L. are shown for results consistent with zero at the 3σ level.

| Isotope | Activity [mBq/kg] |
|---|---|
| $^{228}$Ac | < 1.0 |
| $^{228}$Th | < 1.0 |
| $^{226}$Ra | 5.1 ± 0.4 (stat) ± 2.2 (syst) |
| $^{40}$K | < 16.4 |

When possible configurations of the labels and their contaminants were considered, the minimum and the maximum $^{226}$Ra activities were, as shown in Table 4, 3.4 ± 0.4 (stat) mBq/kg from case 2 and 6.9 ± 0.3 (stat) mBq/kg from case 1. The average result for the $^{226}$Ra activity was 5.1 ± 0.4 (stat) ± 2.2 (syst) mBq/kg. The statistical uncertainty is conservatively selected from the larger of the two cases considered. The systematic uncertainty is dominated by half of the difference in the results from the two cases considered and is presented in Table 4. Additionally, a 10% detection efficiency uncertainty was included in both the label and the powder activity measurements.

We report a positive observation for the $^{226}$Ra activity in spite of the superficial appearance that the result does not pass our stated 3σ criterion. The label geometry errors dominate the systematic uncertainty. This uncertainty accounts for the full range of possible detection efficiencies from different possible configurations of the labels, and as such is not a Gaussian uncertainty.

For application to AMoRE-II, the most crucial goal of this measurement was to constrain the activity of $^{208}$Tl from the $^{228}$Th chain. Only upper limits were found for all activities, except for $^{226}$Ra. These limits were dominated by statistical uncertainties determined by the detector sensitivity, counting time, and limits on sample mass. An array of fourteen HPGe detectors named CAGe, having better-combined sensitivity than CC1, was installed in Y2L [11, 15, 21, 22]. Currently, other powder samples are being assayed with the CAGe to improve measurements of $^{208}$Tl and other activities. Other cosmogenic isotopes, such as $^{88}$Y [23, 24], may be investigated as well.

## ACKNOWLEDGMENT

This work was supported by the Institute for Basic Science (IBS) funded by the Ministry of Science and Technology, Korea (Grant id: IBS-R016-D1).